# Nonsaturating magnetoresistance and nontrivial band topology of type-II Weyl semimetal NbIrTe$_4$

*W. Zhou[1], B. Li[2], C. Q. Xu[1], M. R. van Delft[3], Y. G. Chen[4], X. C. Fan[1], B. Qian[1]\*, N. E. Hussey[3] and Xiaofeng Xu[1]\**

1. Department of Physics and Advanced Functional Materials Lab, Changshu Institute of Technology, Changshu 215500, China
E-mail: njqb@cslg.edu.cn, xiaofeng.xu@cslg.edu.cn
2. Information Physics Research Center, Nanjing University of Posts and Telecommunications, Nanjing 210023, China
3. High Field Magnet Laboratory (HFML-EMFL), Radboud University, 6525ED Nijmegen, Netherlands
4. Department of Physics, Fudan University, Shanghai 200433, China

Weyl semimetals, characterized by nodal points in the bulk and Fermi arc states on the surface, have recently attracted extensive attention due to the potential application on low energy consumption electronic materials. In this report, the thermodynamic and transport properties of a theoretically predicted Weyl semimetal NbIrTe$_4$ is measured in high magnetic fields up to 35 T and low temperatures down to 0.4 K. Remarkably, NbIrTe$_4$ exhibits a nonsaturating transverse magnetoresistance which follows a power-law dependence in $B$. Low-field Hall measurements reveal that hole-like carriers dominate the transport for $T > 80$ K, while the significant enhancement of electron mobilities with lowering $T$ results in a non-negligible contribution from electron-like carriers which is responsible for the observed non-linear Hall resistivity at low $T$. The Shubnikov-de Haas oscillations of the Hall resistivity under high $B$ give the light effective masses of charge carriers and the nontrivial Berry phase associated with Weyl fermions. Further first-principles calculations confirm the existence of 16 Weyl points located at $k_z = 0$, $\pm 0.02$ and $\pm 0.2$ planes in the Brillouin zone.

## 1. Introduction

The recent advent of 3D Dirac/Weyl semimetals is a conceptual breakthrough in the theory and classification of metals.[1–7] Dirac semimetals are featured by Dirac points where the energy momentum dispersions develop linearly along all three momentum directions. In the presence of broken time-reversal symmetry (TRS) or space-inversion symmetry, Dirac semimetals evolve into Weyl semimetals and Dirac points split into pairs of Weyl points due to the lifted spin degeneracy.[8,9] For both the Dirac and Weyl semimetals, due to the coexistence of the conventional charge carriers and relativistic carriers, interesting transport phenomena can often be observed, for example, the nonsaturating magnetoresistance (MR) effect,[10] chiral anomaly,[11,12] and ultrahigh carrier mobilities.[13] Accordingly, transport studies are viewed as an important way to explore the unique scattering mechanism of charge carriers in Dirac/Weyl semimetals. In addition, through transport quantum oscillations, one can get deep insights into the underlying band topology of Dirac/Weyl semimetals.

In general, Weyl semimetals can be classified into two types: the standard type I which possesses a pointlike Fermi surface, and the type II with strongly tilted Weyl cones induced by the broken Lorentz symmetry.[14] The Weyl points for type II Weyl semimetals appear only at the contact of electron and hole pockets. The qualitatively distinct band topology of type II Weyl semimetals can lead to marked differences in physical properties, such as direction restricted chiral anomaly,[15,16] and exotic superconductivity.[17]

To date, several non-centrosymmetric material systems, for example, the transition metal dichalcogenides $T_MX_2$ ($T_M=$ W, Mo, etc; $X=$ Se, Te ),[14,18–20] the diphosphides (Mo, W)$P_2$,[21] LaAlGe,[22] and etc., have been considered to be type II Weyl semimetals, while most of them display complex band structures with multiple Weyl points, which adds difficulties in the conventional transport studies of these materials. Moreover, in many Weyl semimetals, the Weyl points are located deeply below the Fermi level, which often hinders the observation of intrinsic characteristics associated with the relativistic carriers.

Very recently, ternary $MT$Te$_4$ ($M=$ Nb or Ta; $T=$ Ir or Rh) compounds were theoretically predicted as a new series of type II Weyl semimetals,[23,24] and verified experimentally in TaIrTe$_4$.[25–27] In comparison with the previous type II Weyl semimetals, the angle-resolved photoemission spectroscopy (ARPES) data and band calculation of TaIrTe$_4$ suggests the existence of only four Weyl points,[24,26] which is the minimum number of Weyl points allowed for a time-reversal invariant Weyl semimetal. Surface superconductivity was even observed in TaIrTe$_4$, suggesting that TaIrTe$_4$ is a promising platform to explore topological superconductivity.[28] Remarkably, it was experimentally suggested the Weyl cones in TaIrTe$_4$ are located just ∼40-50 meV above the chemical potential,[25] which means one may achieve Weyl points close to the Fermi level via tuning the chemical potential of TaIrTe$_4$, for example through doping, external pressure, or electrostatic gating.

By contrast, the Fermi topology of Nb-based counterpart NbIrTe$_4$ is still unknown. In this paper, we have investigated the underlying band structure of NbIrTe$_4$ by a detailed transport study and the first-principle calculation. The transport of NbIrTe$_4$ displays nonsaturating MR and multiband behavior induced by mixing of the low-mobility hole-like carriers and the high-mobility electron-like carriers. The Shubnikov-de Haas (SdH) oscillations further confirms relatively light effective masses of the carriers and a non-trivial Berry phase. The experimental data also reveal good consistency with our band calculation results.

**2. Results and discussion**

The ternary NbIrTe$_4$ crystallizes in a noncentrosymmetric orthorhombic layered structure with the same space group (*Pmn2$_1$*) as WTe$_2$, as schematically shown in figure 1 (a). In comparison with WTe$_2$, the unit cell along the *b*-axis doubles for the repeated alignments between the two metallic elements Nb and Ir. Along the *a*-axis, alternating Nb and Ir atoms are connected in the form of quasi-one-dimensional zigzag chains. The hybridizations of these chains along the *b*-direction can cause Peierls distortioninduced dimerization as found in many monolayer $T_MX_2$,[14,29] which can result in a variety of novel quantum phenomena such as nontrivial electronic topology and a nonsaturating magnetoresistance (MR) effect. The single-crystal x-ray diffraction (XRD) pattern is shown in figure 1 (b). The sharp XRD peaks indicate the good quality of the samples. The observed peaks can be well indexed with the (002*l*) peaks, which suggests that the crystallographic *c* axis is perfectly perpendicular to the surface facet of the crystal. An optical image of an as-grown crystal was shown in figure 1 (b). The long edge of the strip-like crystal is the crystallographic *a*-axis. In figure 1 (c), the resistivity $\rho_a$ ($I \parallel a$) and $\rho_c$ ($I \parallel c$) curves collected from room temperature down to 2 K are shown. The resistivity $\rho_b$ ($I \parallel b$) value is comparable to $\rho_a$ (data not shown). For $\rho_a$ and $\rho_c$, though similar metallic behaviors are shown, large anisotropy ratios of $\rho_c/\rho_a$ over several tens were found across the measured $T$ range. We have also compared the longitudinal magnetoresistance (LMR) for these two current directions (inset in figure 1 (c)). As seen, LMR for $I \parallel a$ is very weak, while the LMR for $I \parallel c$ reaches ~70% under a field of 9 T. The transverse magnetoresistance (TMR) for $I \parallel a$ is also measured and shown in figure 1 (d). With increasing temperature, the TMR is gradually suppressed. By a Kohler plot, all TMR curves measured at different temperatures can be scaled into a single curve, indicating that Kohler's rule is well obeyed over a large temperature range (inset in figure 1 (d)). Such behaviors have also been observed in $WTe_2$ showing extremely large magnetoresistance (XMR), which was used as evidence excluding the possible existence of a magnetic-field-driven metal-insulator transition or significant contribution of an electronic phase transition to the low-temperature XMR.[30]

The high field TMR ($B \perp I$) and LMR ($B \parallel I$) at $T$=1.4 K are also characterized and shown in figure 2 (a) and (b). For TMR, the MR value follows a power law as a function of $B$ (i.e., MR $\propto B^\alpha$), and there is no tendency of saturation. Such power-law $B$-dependent and nonsaturating TMR has also been observed in $TaIrTe_4$.[25] No negative LMR is observed (Figure 2 (b)). The LMR appears to follow a sublinear $B$-dependence, in contrast to the nonsaturating behavior of TMR. Note that, chiral anomaly in Weyl semimetals can manifest itself in a negative LMR, due to the

Adler-Bell-Jackiw chiral anomaly,[31] which has been demonstrated in many type-I Weyl semimetals.[11,12] For type-II Weyl semimetals, however, the chiral anomaly is hard to be observed due to the tilted nature of the Weyl cones.

To make clear the origin of the nonsaturating TMR behavior, it is important to explore the scattering mechanism of the charge carriers and also the band topology. Accordingly, we performed the Hall and SdH measurements. In figure 2 (c), the field dependence of the Hall resistivity $\rho_{xy}$ measured at different temperatures are shown. At high temperatures ($T > 80$ K), $\rho_{xy}$ is positive and develops linearly with $B$, indicating that single hole-type carriers dominate the transport. The hole density can be simply obtained by $n_h = 1/eR_H$, where $R_H$ is the Hall coefficient defined by the slope in $\rho_{xy}(B)$ curve. For $T \leq 80$ K, the $\rho_{xy}(B)$ curve becomes nonlinear, implying a prominent multiband effect at low temperatures. At the lowest temperatures, $\rho_{xy}$ changes sign from positive to negative, indicating that the electrons was becoming increasingly more mobile. In light of this, we have fitted the $\rho_{xy}(B)$ curves with the single-band model for $T > 80$ K and two-band model for $T \leq 80$ K. In the latter case, the Hall resistivity is described by

$$\rho_{xy} = \frac{B}{e} \frac{(\mu_h^2 n_h - \mu_e^2 n_e) + (\mu_h \mu_e)^2 B^2 (n_h - n_e)}{(\mu_e n_e + \mu_h n_h)^2 + (\mu_h \mu_e)^2 B^2 (n_h - n_e)^2} \quad (1)$$

The carrier density $n$ and mobility $\mu$ are shown in figures 2 ((d) and (e)), respectively. In order to obtain $n_h$ in the high-$T$ single-band region, zero-field resistivity was used. It is apparent that the hole density $n_h$ is higher than the electron density $n_e$, while the mobility of electron ($\mu_e$) grows much faster than that of hole carriers ($\mu_h$). Clearly, these results exclude the electron-hole compensation explanation for the nonsaturating TMR proposed for some type II Weyl semimetals. The significant growth of electron mobility with decreasing $T$ is the direct reason for the evolution of $\rho_{xy}$ with field and temperature. Consistent with the Hall coefficient, the TEP also changes sign with decreasing $T$. The TEP is characterized by a large peak around 10 K, presumably due to the phonon-drag effect as that in Dirac semimetal NiTe$_2$.[32] The heat capacity characterization of the sample shows no evident anomaly below 200 K (Figure 2 (g)). The separation of electronic ($\gamma T$) and phononic ($\beta T^3$) contributions at low temperatures yields $\gamma$ = 2.68 mJ/mol K$^2$, $\beta$ = 1.93 mJ/mol K$^4$, and the Debye temperature $\Theta_D$ ~ 182 K (inset of Figure 2(g)).

To probe the topology of the underlying Fermi surface, we have measured the magnetoresistance and Hall resistivity of NbIrTe$_4$ over a wide field and temperature range. The SdH oscillations from the Hall component and their Fast Fourier transform (FFT) are presented in figure 3 (a) and (c), respectively. Sizeable and beautiful quantum oscillations are observed in the Hall resistivity curve for $B$ over ~10 T and $T$ < 13 K. In total, five frequencies are identified, which are labeled as $F$1, $F$2, $F$3, $F$4, and $F$5 from low to high. From the Onsager relation $F = (\Phi_0/2\pi^2)S_F$ (where $\Phi_0$ is the flux quantum), the extremal Fermi surface cross-sectional area $S_F$ for each frequency can be obtained (see Table I). Since the weak $T$ dependence of the first frequency is quite abnormal, we will ignore further analysis on $F$1. In general, the oscillatory SdH data can be described by the standard Lifshitz-Kosevich (LK) formula:[33,34]

$$\Delta\rho \propto R_T R_D R_S \cos[2\pi(\frac{F}{B} + \frac{1}{2} - \frac{\phi_B}{2\pi} - \delta)] \qquad (2)$$

where $R_T$, $R_D$, $R_S$ are the thermal damping factor, Dingle damping term and a spin-related damping term, respectively, $\phi_B$ is the Berry phase, and $\delta$ is an additional phase shift determined by the dimensionality of the Fermi surface, that is $\delta=0$ ($\delta=\pm 1/8$) for 2D (3D) Fermi surfaces. Therein, $R_T = \frac{2\pi^2 k_B T m^*/eB\hbar}{\sinh(2\pi^2 k_B T m^*/eB\hbar)}$, $R_D = \exp(2\pi^2 k_B T_D m^*/eB\hbar)$, and $R_S = \cos(\pi g m^*/2m_e)$, where $m^*$ is the effective electron mass, $m_e$ is free electron mass, $k_B$ is the Boltzmann constant, and $T_D$ is the Dingle temperature. From the LK formula, the effective mass $m^*$ can be obtained from the fit of the temperature dependence of its corresponding FFT amplitude, as shown in Figure 3 (d). The best fits obtained effective masses are 0.29 ±0.02 $m_e$ (F2), 0.42 ±0.02 $m_e$ (F3), 0.33±0.02 $m_e$ (F4), and 0.60±0.02 $m_e$ (F5). To analyze the expected "zero mode" which corresponds to a nontrivial Berry phase for Weyl semimetals, we have constructed the Landau level (LL) index fan diagram for each frequency. Taking $F2$ as an example (see figure 3 (e) and (f)), we first calculate the Hall conductivity $\sigma_{xy}$ through the longitudinal resistivity $\rho_{xx}$ and the Hall resistivity $\rho_{xy}$ using $\sigma_{xy} = \rho_{xy}/(\rho_{xy}^2 + \rho_{xx}^2)$, and get $\Delta\sigma_{xy}$ by substracting the background. Then, we adopt the valley (and peak) of -$\Delta\sigma_{xy}$ as the index field $B_n$, as done in ref. [35]. Further, $B_n$ is plotted as a function of $n+1/4$, where $n$ is the integer LL index. It is well known that the Berry phase is zero for a parabolic energy dispersion and π for a linear energy dispersion. As obtained from the linear fit in figure 3 (f), the slope of the LL indices is 145.4 T for $F2$, in good agreement with the frequencies obtained from the FFT analysis, and the intercept $n_0$ is -0.47 ±0.02. The corresponding Berry phase is calculated by $\phi_B = 2\pi|-0.47 + \delta|$. The values of $\phi_B$ for $F3$, $F4$, and $F5$ obtained from the same procedure are listed in Table I.

**Table I** Parameters obtained from SdH oscillations of a NbIrTe$_4$ single crystal, in which $S_F$ is the extremal Fermi-surface cross-sectional area calculated from the Onsager relation, $m^*$ and $m_e$ are the effective electron mass and the bare electron mass, respectively, and $\phi_B$ is the Berry phase.

| Parameter | SdH ($B \parallel c$) NbIrTe$_4$ | | | |
|---|---|---|---|---|
| | $F2$ | $F3$ | $F4$ | $F5$ |
| Frequency (T) | 145.5 | 290 | 418.5 | 582.1 |
| $S_F$ ($10^{-2}$ Å$^{-2}$) | 1.39 | 2.77 | 3.99 | 5.56 |
| $m^*/m_e$ | 0.29±0.02 | 0.42±0.02 | 0.33±0.02 | 0.60±0.02 |
| $\phi_B$ ($\delta = -1/8$) | 1.2π | 0.53π | 0.03π | 0.55π |
| $\phi_B$ ($\delta = 0$) | 0.94π | 0.28π | 0.28π | 0.80π |
| $\phi_B$ ($\delta = 1/8$) | 0.69π | 0.03π | 0.53π | 1.05π |

The angle dependence of the SdH oscillations is also measured and shown in figure 4 (a). Based on the FFT spectrums at different angles, we extract the angle dependency of each frequency (solid symbols in figure 4 (c)). It is noticed, all frequencies are present near $\theta = 90°$ (or $B \parallel c$-axis), and they almost all disappear for $\theta < 50°$.

To further identify the contributing Fermi pocket to each oscillation frequency, we performed first-principle calculation. The experimentally obtained angular frequency values can generally coincide with those of band3 and band4 from the first-principles calculations (see figure 4(c)). The calculated bulk band structure of NbIrTe$_4$ with SOC along high symmetry path is shown in figure 5 (a). There are four bands crossing the Fermi level. Two nested hole-like bands (band1 and band2) are plotted in magenta and blue color while two nested electron-like bands (band3 and band4) are shown in green and red color. Spin-orbit coupling splits the bands such that the hole pocket band2 (figure 5 (d)) is contained in the hole pocket band1 (figure 5 (c)). The electron pockets showing a large anisotropy with open orbits along the $k_z$ direction, also split into the inner electron pocket band4 (figure 5 (f)) and the outer electron pocket band3 (figure 5 (e)), which exhibits a noticeable two-dimensional character in the $k_z$ direction. We also carried out the bulk Fermi surface at $k_z = 0$ and surface state spectral at (001) slab as shown in figure 5 (b). We can clearly see the extra surfaces states (bright red lines) appear on both sides of the bulk states. The absence of inversion symmetry introduces a topological phase transition of NbIrTe$_4$ into Weyl semimetal phase. We performed a scan of bulk band structure in full Brillouin zone to locate the Weyl points and found that NbIrTe$_4$ has 16 Weyl points, with 8 nodes locating at $k_z = \pm 0.2$ plane ($\pm 0.04, \pm 0.19, \pm 0.2$), 4 nodes at $k_z = \pm 0.02$ plane ($\pm 0.07$, 0.06, 0.02), ($\pm 0.07$, -0.06, -0.02), and 4 nodes at $k_z = 0$ plane ($\pm 0.14, \pm 0.21, 0$).

## 3. Conclusion

In summary, we have performed a detailed investigation of the transport properties of a predicted Weyl semimetal NbIrTe$_4$. Anisotropic magnetoresistance behaviors are found for fields and currents applied along different crystallographic axis, and the TMR for $I \parallel a$ exhibits power law dependence on $B$ up to 35 T. Temperature dependent Hall measurements suggest a significant enhancement of charge carrier mobilities of electrons with lowering $T$ which may be associated with the 'relativistic' carriers in Weyl systems. High field Hall resistivity measurements reveal clear SdH oscillations from which the light effective masses of charge carriers and the nontrivial Berry phase are derived. The band structure calculations further confirm the existence of four Weyl cones. Overall, our data consistently suggests that NbIrTe$_4$ provides a new platform to study the nontrivial physics belonging to 'relativistic' carriers.

*Note added*. In the proof of this work, we recently noticed another work investigating the band topology of NbIrTe$_4$ through quantum oscillations of the magneto-resistivity component[36].

## 4. Experimental Section

Single crystals of NbIrTe$_4$ were grown out of excess Te by the flux method. High purity Nb, Ir, Te elements were mixed with a molar ratio of 1: 1: 20. The mixtures were put into alumina crucibles, sealed in vacuumed quartz tubes, and heated to 1273 K in a hightemperature box furnace. After keeping at 1273 K for two days, the samples were slowly cooled down to 700 K. Finally, long and stripe-like single crystals which can be easily exfoliated along the *ab*-plane were harvested.

Single-crystal x-ray diffraction (XRD) measurements were performed at room temperature using a diffractometer with Cu *Ka* radiation and a graphite monochromator. The elemental compositions of a set of single crystals were measured by the energy dispersive X-ray spectroscopy (EDX), proving that the stoichiometry ratio for Nb:Ir:Te is very close to 1:1:4. Low-field Hall-resistivity/MR, angle-dependent MR, and specific heat data were all collected on a Quantum Design (QD) physical property measurement system (PPMS). The standard 4-probe method was employed for these electrical measurements. The Shubnikov-de Haas (SdH) oscillations from the Hall component in high magnetic field were acquired with a resistive Bitter magnet at the High Field Magnet Laboratory (HFML) (Nijmegen) with a maximum field of 35 T. The Hall effect was performed by reversing the field direction and antisymmetrizing the data. For the thermoelectric power (TEP) measurements, a modified steady-state method was used in which a temperature gradient, measured using a constantan-chromel differential thermocouple, was set up across the sample via a chip heater attached to one end of the sample. The TEP voltage was read out by a nanovoltmeter K2182A from Keithley Instruments.

Based on our experimental crystal structure, we carried out first-principles calculations for band structure and Fermi surfaces. The electronic structure calculations with high accuracy were performed using the full-potential linearized augmented plane wave (FP-LAPW) method implemented in the WIEN2K code.[37] The generalized gradient approximation (GGA) is applied to the exchange-correlation potential calculation.[38] The muffin tin radii are chosen to be 2.41 a.u. for both Nb and Te, 2.5 a.u. for Ir. The plane-wave cutoff is defined by $RK_{max} = 7.0$, where R is the minimum LAPW sphere radius and $K_{max}$ is the plane-wave vector cutoff. In order to get accurate electronic Fermi surfaces, we used a 84× 26× 24 k-mesh with about 50000 K-points to calculate the eigen energies which are accurate enough to evaluate the result. Spin-orbit coupling was included in the calculations. To obtain the surface state properties, we constructed a tight-binding model based on maximally localized Wannier functions.[39] The surface state spectrum (SSS) were calculated with the surface Green's function methods as implemented in WannierTools.[40] We use Nb $d$, Ir $d$, and Te $p$ orbitals to construct Wannier functions.

**Acknowledgements**

This work is sponsored by the National Natural Science Foundation of China (Grant No. 11704047, No. U1732162, No.11674054, No. U1832147), by Natural Science Foundation of Jiangsu Educational Department (Grant No. 15KJA430001) and six-talent peak of Jiangsu Province (Grants No. 2017-XCL-001), by the Foundation of Jiangsu Science and Technology Department (Grant No. BA2016041), by the science and technology development plan project in Suzhou (SZS201710). We acknowledge the support of the HFML, member of the European Magnetic Field Laboratory (EMFL). X. X. would also like to acknowledge the financial support from an open program from Wuhan National High Magnetic Field Center (2015KF15). W. Zhou, B. Li, and C. Q. Xu contributed equally to this work.

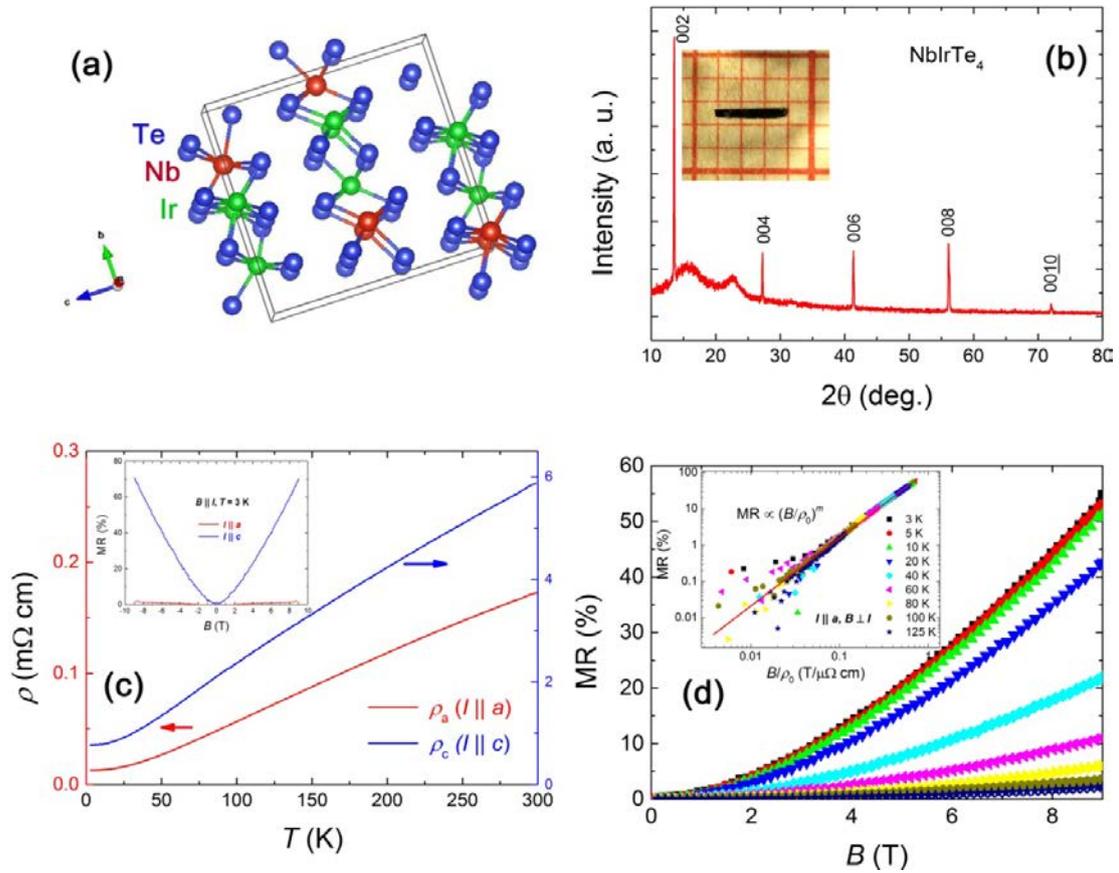

**Figure 1** (a) Atomic crystal structure of NbIrTe$_4$. (b) Single-crystal diffraction pattern (XRD) of NbIrTe$_4$. The inset is an optical image of the as-grown crystal. (c) Temperature dependence of resistivity $\rho_a$ and $\rho_c$ with the current applied along the *a*-axis and *c*-axis, respectively. Inset: the longitudinal magnetoresistance (LMR) for $\rho_a$ and $\rho_c$ with field *B* parallel to the current directions. MR in figure is defined as $100\% * \frac{\rho(B)-\rho_0}{\rho_0}$ in which $\rho_0$ is the resistivity at zero field. (d) Field dependence of the transverse magnetoresistance (TMR) for $\rho_a$ at different temperatures. Inset: Kohler's scaling of TMR. The red solid line is a fit based on MR $\propto \left(\frac{B}{\rho_0}\right)^m$ with *m*~ 1.71.

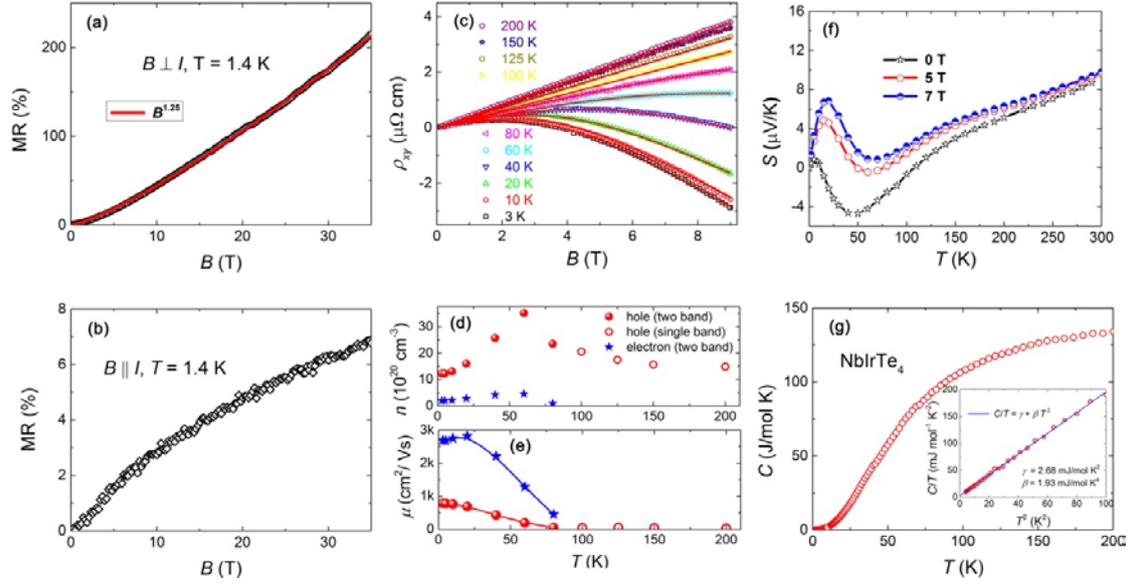

**Figure 2** (a) and (b) TMR ($B \perp I$) and LMR ($B \parallel I$) of $\rho_a$ measured with field $B$ up to 35 T and $T$=1.4 K. (c) The Hall resistivity at different temperatures. The red solid lines are fits based on two-band model for $T \leq 80$ K and single-band model for $T > 80$ K. (d) and (e) Temperature dependence of the carrier density $n$ and the mobility $\mu$ extracted from the above fits. (f) Temperature dependence of the thermoelectric power $S$ at $B = 0, 5, 7$ T. (g) Temperature dependence of heat capacity $C$ of NbIrTe$_4$ single crystal. Inset shows a fit to $C(T) = \gamma T + \beta T^3$ in low $T$ region.

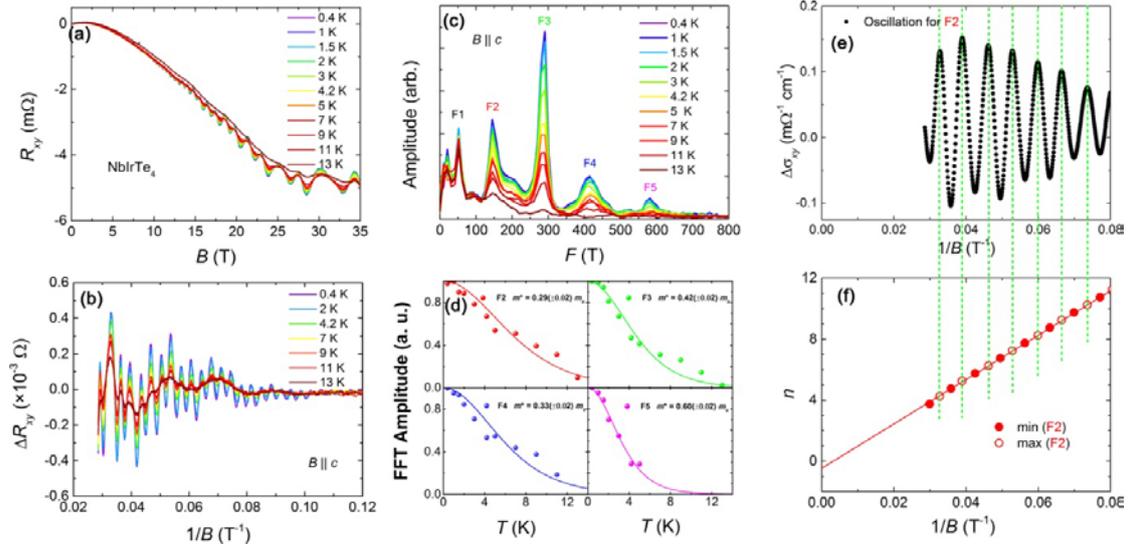

**Figure 3** SdH quantum oscillation of NbIrTe$_4$ (a) Original Hall signals ($\rho_{xy}$) collected at different temperatures. (b) Oscillatory Hall component $\Delta R_{xy}$ after substracting the background plotted as a function of $B^{-1}$. (c) The corresponding FFT spectrum from the SdH oscillation. (d) The FFT amplitude as a function of temperature and the fits to $R_T$ to determine the effective masses for $F2$, $F3$, $F4$, and $F5$. (e) Field dependence of the oscillatory Hall conductivities $\Delta\sigma_{xy}$ ($T = 1.5$K) of $F2$, which is obtained via FFT band filter. (f) Landau level index fan diagram for $F2$, which is constructed from the peak (maximum) and valley (minimum) on the 1.5 K oscillatory $\Delta\sigma_{xy}$ in panel (e).

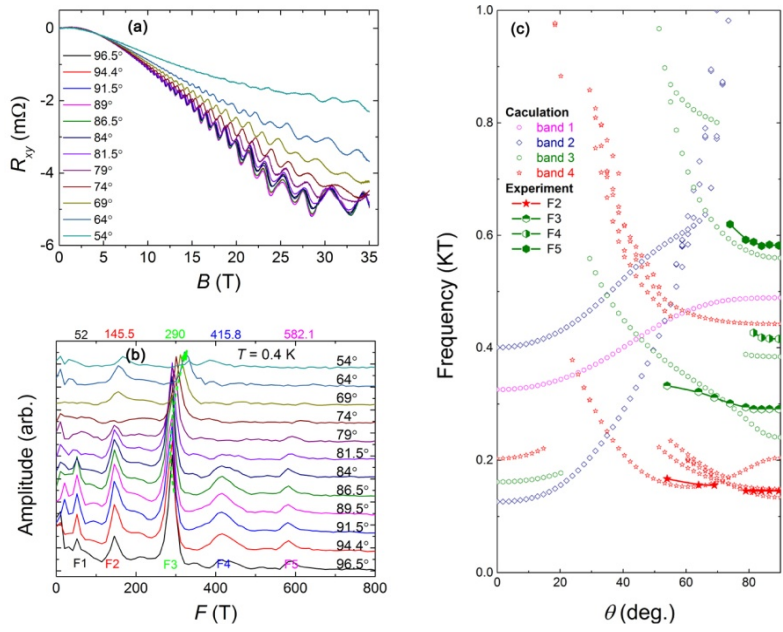

**Figure 4** (a) SdH quantum oscillations at different angles ($T = 0.4$ K). Angle $\theta = 90°$ ($0°$) corresponds to $B \parallel c$-axis ($b$-axis).(b) Corresponding FFT spectrums at different angles. (c) The angular dependence of FFT frequency. The solid symbols are experimental data determined from the FFT spectrum of panel (b). The open symbols are data based on the theoretical calculation. In panel (c), bands are color-coded, i.e., each calculation band is plotted with the same color as its corresponding experimental band.

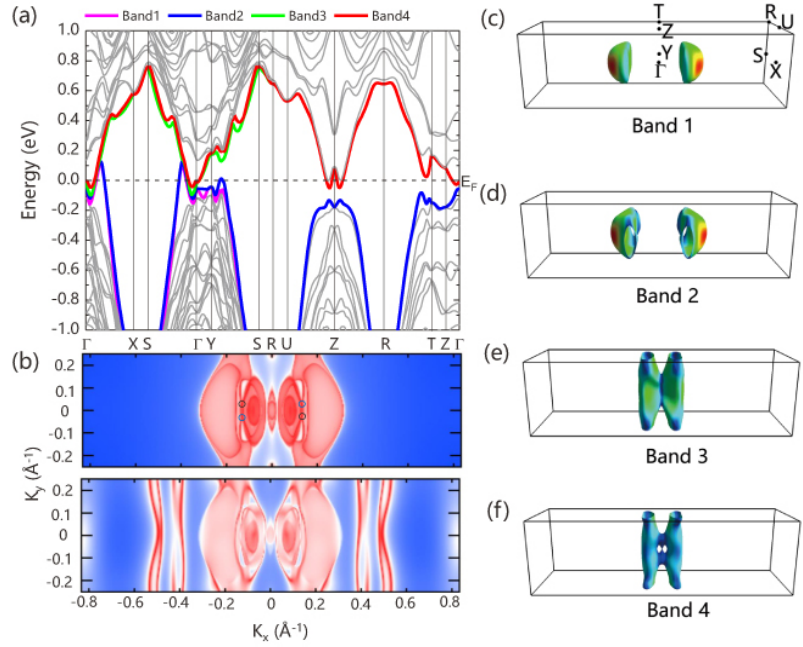

**Figure 5** (a) Calculated band structure along Γ -X -S -Γ -Y -S -R-U - Z - R - T - Z -Γ. Four bands crossing the Fermi level are marked by different colors. (b) top: Fermi surfaces at $k_z = 0$ for bulk NbIrTe$_4$, four Weyl nodes on this plane are illustrated in circles; bottom: surface state spectral at (001) slab. (c-f) 3D Fermi surfaces for each band, shaded by Fermi velocity [darker (blue) is low velocity].